# Residuated Park Theories

Z. Ésik*
Dept. of Computer Science
University of Szeged
6720 Szeged
Árpád tér 2

**Abstract**

When $L$ is a complete lattice, the collection $\mathbf{Mon}_L$ of all monotone functions $L^p \to L^n$, $n, p \geq 0$, forms a Lawvere theory. We enrich this Lawvere theory with the binary supremum operation $\vee$, an operation of (left) residuation $\Leftarrow$ and the parameterized least fixed point operation $^\dagger$. We exhibit a system of *equational* axioms which is sound and proves all valid equations of the theories $\mathbf{Mon}_L$ involving only the theory operations, $\vee$ and $^\dagger$, i.e., all valid equations not involving residuation. We also present an alternative axiomatization, where $^\dagger$ is replaced by a star operation, and provide an application to regular tree languages.

## 1 Introduction

The semantics of recursion is usually described by fixed points of functions, functors, or other constructors. Least fixed points of monotone or continuous functions on cpo's or complete lattices have been widely used to give semantics to functional programs and various programming logics. The parameterized least fixed point operation $^\dagger$, in conjunction with function composition and the cartesian operations (or Lawvere theory operations) satisfies several nontrivial equations, such as the well-known De Bakker-Scott-Bekić equation, [5, 2]. One would naturally like to have a complete description of *all* valid (in)equations in the form of a system of axioms. Such a complete description is given by the axioms of Iteration Theories [4, 10], see also [20] and [1]. The equational axioms of iteration theories can be divided into two parts, the relatively simple "Conway equations", and a complicated equation associated with each finite (simple) group. In contrast, by allowing quasi-equations (or implications) in addition to equations, very simple axiomatizations can be given. For example, the system given in [9, 12] involves two simple equations, the fixed point equation and the parameter equation, and an implication, the least pre-fixed point rule (also known as the fixed point induction axiom). In the ordered setting, these axioms are sound and complete with respect to all valid equations and inequations of the least fixed point operation on monotone or continuous functions on complete lattices, or cpo's.

*Partially supported by the project TÁMOP-4.2.1/B-09/1/KONV-2010-0005 "Creating the Center of Excellence at the University of Szeged", supported by the European Union and co-financed by the European Regional Fund, and the National Foundation of Hungary for Scientific Research, grant no. K 75249.

The Lawvere theory of monotone (or continuous) functions on a complete lattice can also be equipped with the pointwise binary supremum operation ∨. It was shown in [11] that the least fixed point rule and a few simple (in)equations are complete for the equational properties of the least fixed point operation in conjunction with the theory operations and the operation ∨ in these theories of functions. (These operations may be called the "rational operations".)

The contribution of the present paper is that in the presence of (one-sided) residuation, which can be defined in the theories of monotone or continuous functions on complete lattices, the least fixed point rule can be replaced by pure equational axioms. We present a simple system of equations, involving the theory operations, $^\dagger$, ∨ and residuation, which is sound in all theories of monotone or continuous functions on complete lattices, and which is complete for those (in)equations involving only the rational operations (i.e., which do *not* involve residuation). We also present an alternative axiomatization, where $^\dagger$ is replaced by a star operation, and provide an application of the completeness results to regular tree languages.

The program carried out in this paper stems on one side from [16] and [8, 11, 12], and from [18, 19] on the other side. Kozen [16] proved the important result that a finite set of simple equations together with the least pre-fixed point rule (for the star operation) and its dual capture the equational theory of regular languages and idempotent continuous semirings enriched with Kleene star. Subsequently, Pratt [18] showed the beautiful theorem that by adding (two-sided) residuation to the Kleene algebras of Kozen, it is possible to obtain a finitely based equational theory which is a conservative extension of the equational theory of Kleene algebras, and thus complete for regular languages and for idempotent continuous semirings equipped with the operation of Kleene star. An interesting generalization of Pratt's ideas is given in [19].

## 2 The contribution

When $A$ is any set, the *Lawvere theory of functions* $\mathbf{F}_A$ is a small category whose objects are the natural numbers and whose morphisms $n \to p$, $n, p \geq 0$ are the functions $A^p \to A^n$ (note the reversal of the arrow). The *composite* of $f : n \to p$ and $g : p \to q$ is their function composition, denoted $f \cdot g : n \to q$, so that $f \cdot g : A^q \to A^n$. In the Lawvere theory $\mathbf{F}_A$, the $i$th *distinguished coproduct injection* $i_n : 1 \to n$, $i \in [n] = \{1, \ldots, n\}$, $n \geq 0$, is the $i$th projection function $A^n \to A$. By the coproduct property, for any sequence $f_1, \ldots, f_n : 1 \to p$ there is a unique $f : n \to p$ with $i_n \cdot f = f_i$, for all $i \in [n]$. We denote this unique function by $\langle f_1, \ldots, f_n \rangle$ and call it the *tupling* of the $f_i$, $i \in [n] = \{1, \ldots, n\}$. When $n = 0$, we also write $0_p$. Note that $\mathbf{1}_n = \langle 1_n, \ldots, n_n \rangle$ is the identity function on $A^n$, for all $n \geq 0$. The composition and tupling operations and the constants $i_n$ satisfy the following equations:

$$\begin{aligned}
(f \cdot g) \cdot h &= f \cdot (g \cdot h), \quad f : m \to n,\ g : n \to p,\ h : p \to q \\
\mathbf{1}_n \cdot f &= f = f \cdot \mathbf{1}_p, \quad f : n \to p \\
i_n \cdot \langle f_1, \ldots, f_n \rangle &= f_i, \quad f_1, \ldots, f_n : 1 \to p,\ i \in [n],\ n \geq 0 \\
\langle 1_n \cdot f, \ldots, n_n \cdot f \rangle &= f, \quad f : n \to p \\
\mathbf{1}_1 &= 1_1.
\end{aligned}$$

Here, the first two equations express that $\mathbf{F}_A$ is a category, and the 3rd and 4th equations assert that each object $n$ is the $n$-fold coproduct of object 1 with itself. In particular, 0 is an initial object. The last equation asserts that the distinguished coproduct injection $1 \to 1$ is in fact the identity morphism $\mathbf{1}_1 = \langle 1_1 \rangle$.

**Definition 2.1** *A* Lawvere theory *is any small (cartesian) category whose objects are the natural numbers such that each object $n$ is the $n$-fold coproduct of object $1$ with itself.*

We further assume that each theory comes with chosen coproduct injections $i_n : 1 \to n$, $i \in [n]$, $n \geq 0$, and that $1_1$ is the identity morphism $1 \to 1$. (This implies that $\langle f \rangle = f$ for all $f : 1 \to p$.) It is well-known that each Lawvere theory can be embedded by an object and coproduct preserving functor in a theory $\mathbf{F}_A$, for some set $A$. Thus, each theory can be faithfully represented as a theory of functions.

The above elementary categorical definition of theories may be complemented with an algebraic one. Each theory $T$ may be seen as a many-sorted algebra whose set of sorts is the set $\mathbb{N} \times \mathbb{N}$ of all ordered pairs of nonnegative integers. The carrier of sort $(n, p)$, denoted $T(n, p)$, is called the hom-set of morphisms $n \to p$. The operations and constants are composition and tupling, and the distinguished morphisms $i_n : 1 \to n$, $i \in [n]$, $n \geq 0$. An $\mathbb{N} \times \mathbb{N}$-sorted algebra equipped with these operations and constants is a theory if it satisfies the above equations, cf. e.g., [4]. (Our notation for the theory operations and constants originates from [14].) A morphism of theories is just a morphism of $\mathbb{N} \times \mathbb{N}$-sorted algebras, preserving all the operations and constants, or equivalently, a functor preserving objects and distinguished coproduct injections.

When $\rho$ is a function $[n] \to [p]$ in a theory $T$, there is an associated *base morphism* $n \to p$, defined as the tupling
$$\langle (1\rho)_p, \ldots, (n\rho)_p \rangle.$$
In particular, the identity morphism $\mathbf{1}_n$ is the one associated with the identity function $[n] \to [n]$, and $0_p$ is the base morphism associated with the empty function $[0] \to [p]$.

In any theory $T$, we define the *pairing* $\langle f, g \rangle$ of morphisms $f : n \to p$ and $g : m \to p$ to be the unique morphisms $h : n + m \to p$ with $i_{n+m} \cdot h = i_n \cdot f$ if $i \in [n]$; and $i_{n+m} \cdot h = j_m \cdot g$ where $j = i - n$, if $i \in [n+m] \setminus [n]$. Also, for any $f : n \to p$ and $g : m \to q$ we define the *separated sum* $f \oplus g : n + m \to p + q$ as $\langle f \cdot \kappa, g \cdot \lambda \rangle$, where $\kappa : p \to p + q$ and $\lambda : q \to p + q$ are the base morphisms corresponding to the inclusion $[p] \to [p + q]$ and translated inclusion $[q] \to [p + q]$. Note that we have $\kappa = \mathbf{1}_p \oplus 0_q$ and $\lambda = 0_p \oplus \mathbf{1}_q$. The pairing and separated sum operations are associative and $f \oplus 0_0 = 0_0 \oplus f = f$, for all $f : n \to p$. Also, $\langle f, g \rangle \cdot h = \langle f \cdot h, g \cdot h \rangle$, $(f \oplus g) \cdot \langle h, k \rangle = \langle f \cdot h, g \cdot k \rangle$ and $(f \oplus g) \cdot (h \oplus k) = (f \cdot h) \oplus (g \cdot k)$ for all appropriate morphisms $f, g, h, k$.

In the theory $\mathbf{F}_A$, for any $f : n \to p$ and $g : m \to p$, $\langle f, g \rangle : n + m \to p$ is the function $A^p \to A^{n+m}$ mapping $x \in A^p$ to $(f(x), g(x))$. And if $f : n \to p$ and $g : m \to q$, then $f \oplus g : A^{p+q} \to A^{n+m}$ maps any $(x, y) \in A^{p+q}$ with $x \in A^p$ and $y \in A^q$ to $(f(x), g(y))$.

When $A$ is a set with structure, one usually considers a subtheory of $\mathbf{F}_A$ whose morphisms preserve certain features of the structure. For example, let $L = (L, \leq)$ be a complete lattice, so that $L^n$ is also a complete lattice for each $n \geq 0$. Then we can form the theory $\mathbf{Mon}_L$ whose morphisms are the monotone (or order preserving) functions in $\mathbf{F}_L$. Just as $\mathbf{F}_L$, the theory $\mathbf{Mon}_L$ comes with the pointwise order: when $f, g : n \to p$, $f \leq g$ iff $f(x) \leq g(x)$ for each $x \in L^p$. Each hom-set is also a complete lattice: the supremum $f = \bigvee_{i \in I} f_i$ of any family of morphisms $f_i : n \to p$, $i \in I$ exists and is computed pointwise: $f(x) = \bigvee_{i \in I} f_i(x)$ for all $x \in L^p$. The operation of composition is monotone in both arguments, and preserves arbitrary suprema in its first argument:

$$(\bigvee_{i \in I} f_i) \cdot g = \bigvee_{i \in I} (f_i \cdot g)$$

for all $f_i : n \to p, i \in I$ and $g : p \to q$, where $I$ is any index set.

We will consider the theory $\mathbf{Mon}_L$ of monotone functions on a complete lattice $L$ with three more operations. The first one is the operation of *binary supremum* $\vee$, defined on each hom-set $\mathbf{Mon}_L(n,p)$. (This operation in turn determines the order structure by $f \leq g$ iff $f \vee g = g$, for all $f, g : n \to p$.) The second operation is a weak left inverse of composition. For any fixed $g : p \to q$ and for any $n$, since right composition with $g$,

$$\begin{aligned}\mathbf{Mon}_L(n,p) &\to \mathbf{Mon}_L(n,q) \\ f : n \to p &\mapsto f \cdot g : n \to q\end{aligned}$$

preserves arbitrary suprema, for any $h : n \to q$ there is a greatest $f_0 : n \to p$ with $f_0 \cdot g \leq h$. We denote this morphism (function) $f_0$ by $h\Leftarrow g$, and call $\Leftarrow$ the (left) *residuation* operation.

The last operation is the *least fixed point operation*. When $f : n \to n+p$ in $\mathbf{Mon}_L$, $f$ is a monotone function $L^{n+p} \to L^n$. By the Knaster-Tarski fixed point theorem, for each $y \in L^p$, the function $f(-, y)$ over $L^n$ has a *least (pre-)fixed point* $f^\dagger(y)$. It has the property that

$$f(f^\dagger(y), y) \leq f^\dagger(y),$$

i.e., $f^\dagger(y)$ is a pre-fixed point of the function $f(-, y)$, and if $x^n \in L^n$ is any other pre-fixed point, i.e., $f(x,y) \leq x$, then $f^\dagger(y) \leq x$. It is well-known that $f^\dagger(y)$ is in fact a fixed point, i.e., $f(f^\dagger(y), y) = f^\dagger(y)$. It is also well-known that $f^\dagger$, as a function of $y$ is also monotone, so that it is a morphism $n \to p$ in $\mathbf{Mon}_L$. Thus, $\mathbf{Mon}_L$ is equipped with the (parameterized) least (pre-)fixed point operation $\dagger$ taking a morphism $f : n \to n+p$ to a morphism $f^\dagger : n \to p$, for all $n, p \geq 0$. Note that $(\mathbf{1}_n \oplus 0_p)^\dagger$ is the least morphism $n \to p$, i.e., the function that maps any $x \in L^p$ to the least element $(\bot, \ldots, \bot)$ of $L^n$, where $\bot$ is the least element of $L$.

One of the main technical contributions of this paper is the following result:

**Theorem 2.2** *The following inequations and equations hold in all theories $\mathbf{Mon}_L$, where $L$ is any complete lattice.*

$$\begin{aligned}(f \vee g) \vee h &= f \vee (g \vee h), \quad f, g, h : n \to p & (1) \\ f \vee g &= g \vee f, \quad f, g : n \to p & (2) \\ f \vee f &= f, \quad f : n \to p & (3) \\ i_n \cdot (f \vee g) &\leq (i_n \cdot f) \vee (i_n \cdot g), \quad f, g : n \to p, \ i \in [n] & (4) \\ f \cdot g &\leq (f \vee f') \cdot (g \vee g'), \quad f, f' : n \to p, \ g, g' : p \to q & (5) \\ (h \Leftarrow g) \cdot g &\leq h, \quad g : p \to q, \ h : n \to q & (6) \\ f &\leq (f \cdot g) \Leftarrow g, \quad f : n \to p, \ g : p \to q & (7) \\ h \Leftarrow g &\leq (h \vee h') \Leftarrow g, \quad g : p \to q, \ h, h' : n \to q & (8) \\ f^\dagger &\leq (f \vee g)^\dagger, \quad f, g : n \to n+p & (9) \\ f \cdot \langle f^\dagger, \mathbf{1}_p \rangle &\leq f^\dagger, \quad f : n \to n+p & (10) \\ f^\dagger \cdot g &\leq (f \cdot (\mathbf{1}_n \oplus g))^\dagger, \quad f : n \to n+p, \ g : p \to q & (11) \\ (g \Leftarrow \langle g, \mathbf{1}_p \rangle)^\dagger &\leq g, \quad g : n \to p & (12)\end{aligned}$$

*Moreover, if an inequation between terms involving the theory operations and constants and the operations $\vee$ and $^\dagger$ holds in all theories $\mathbf{Mon}_L$, where $L$ is a complete lattice, then it is provable from the above system using standard (many-sorted) equational logic.*

It is clear that the equation

$$i_n \cdot (f \vee g) \;=\; (i_n \cdot f) \vee (i_n \cdot g), \quad f, g : n \to p, \; i \in [n] \tag{13}$$

is a consequence of the above axioms. For each $n, p \geq 0$, let $\bot_{n,p}$ denote $(\mathbf{1}_n \oplus 0_p)^\dagger$. We will show that the following equations are also consequences of the axioms.

$$\begin{align}
(f \vee g) \cdot h &= (f \cdot h) \vee (g \cdot h), \quad f, g : n \to p, \; h : p \to q \tag{14} \\
f \vee \bot_{n,p} &= f, \quad f : n \to p \tag{15} \\
\bot_{n,p} \cdot g &= \bot_{n,q}, \quad g : p \to q. \tag{16}
\end{align}$$

Thus, $\bot_{n,p} = \langle \bot_{1,p}, \ldots, \bot_{1,p} \rangle$ and $\bot_{1,p} = \bot_{1,0} \cdot 0_p = \bot \cdot 0_p$ for all $n, p \geq 0$. Note that (13) amounts to saying that the operation $\vee$ on morphisms $n \to p$ determines and is determined by the same operation on morphisms $1 \to p$. Clearly, (5) asserts that composition is monotone in both arguments, while (8) asserts that $\Leftarrow$ is monotone in its first argument. By (9), $^\dagger$ is also monotone.

The rest of the paper is organized as follows. In Section 3, we define Park theories and semilattice ordered Park theories which are certain ordered or semilattice ordered theories equipped with the least (pre-)fixed point operation as the dagger operation, and recall some completeness results from [8, 11]. Then, in Section 4, we define residuated ordered and residuated semilattice ordered theories and provide several examples. In Section 5 we combine residuation and the least (pre-)fixed point operation and define residuated Park theories and residuated semilattice ordered Park theories that are the models of the axioms of Theorem 2.2. This section contains the proof of Theorem 2.2 and the proof of a related result. In Section 6, we replace the dagger operation by a star operation and provide several equivalence results. In Section 7, we provide an application to regular tree languages.

## 3 Park theories

In this section we define Park theories and semilattice ordered Park theories as certain ordered theories equipped with the least fixed point operation as the dagger operation. After giving several examples, we recall some completeness results from [11, 8].

A *semilattice ordered theory* is a theory $T$ equipped with an operation $\vee$,

$$f, g : n \to p \;\mapsto\; f \vee g : n \to p, \quad n, p \geq 0,$$

satisfying (1) – (5), where by definition $f \leq g$ for $f, g : n \to p$ iff $f \vee g = g$. A morphism of semilattice ordered theories is a theory morphism that preserves $\vee$. Clearly, every semilattice ordered theory is an *ordered theory*, since when $f, g : n \to p$, then by (13) we have $f \leq g$ iff $i_n \cdot f \leq i_n \cdot g$ for all $i \in [n]$, moreover,

$$f \leq f' \; \& \; g \leq g' \;\Rightarrow\; f \cdot g \leq f' \cdot g', \quad f, f' : n \to p, \; g, g' : p \to q. \tag{17}$$

Also, any morphism of semilattice ordered theories preserves $\leq$ and is thus a morphism of ordered theories.

A **Park theory** [8] is an ordered theory $T$ equipped with a *dagger operation* $\dagger : T(n, n+p) \to T(n,p)$, $n,p \geq 0$ satisfying the *fixed point inequation* (10), the *parameter inequation* (11), and the *least pre-fixed point rule*:

$$f \cdot \langle g, \mathbf{1}_p \rangle \leq g \;\Rightarrow\; f^\dagger \leq g, \quad f : n \to n+p, \; g : p \to q. \tag{18}$$

A morphism of Park theories is an ordered theory morphism that preserves the dagger operation.

Suppose that $T$ is a Park theory. It is known that the *fixed point equation* (19) and the *parameter equation* (20) hold in $T$:

$$f \cdot \langle f^\dagger, \mathbf{1}_p \rangle = f^\dagger, \quad f : n \to n+p \tag{19}$$
$$f^\dagger \cdot g = (f \cdot (\mathbf{1}_n \oplus g))^\dagger, \quad f : n \to n+p, \; g : p \to q \tag{20}$$

It is also known that for each $n,p \geq 0$, $\bot_{n,p} = (\mathbf{1}_n \oplus 0_p)^\dagger$ is the least morphism $n \to p$ in $T$ so that (16) holds. Moreover, $i_n \cdot \bot_{n,p} = \bot_{1,p}$ for all $n,p \geq 0$ and $i \in [n]$, and $\dagger$ is a monotone operation:

$$f \leq g \;\Rightarrow\; f^\dagger \leq g^\dagger, \quad f,g : n \to n+p. \tag{21}$$

A **semilattice ordered Park theory** is a Park theory that is a semilattice ordered theory satisfying (14), or

$$(f \vee g) \cdot h \;\leq\; (f \cdot h) \vee (g \cdot h), \quad \text{for all } f,g : n \to p, \; h : p \to q. \tag{22}$$

A morphism of semilattice ordered Park theories is both a Park theory morphism and a semilattice ordered theory morphism.

It is clear that an ordered theory is an ordered Park theory in at most one way, since in an ordered Park theory, for any morphism $f : n \to n+p$, $f^\dagger$ is the least morphism $g : n \to p$ with $f \cdot \langle g, \mathbf{1}_p \rangle \leq g$, or $f \cdot \langle g, \mathbf{1}_p \rangle \leq g$. We call this dagger operation the least pre-fixed point operation. Examples of ordered Park theories include the theories $\mathbf{Mon}_P$ and $\mathbf{Cont}_P$ of monotone and continuous functions over a cpo $P$. In these theories, a morphism $n \to p$ is a monotone or continuous function $P^p \to P^n$. Both theories are ordered by the pointwise order and are equipped with the least fixed point operation as the dagger operation. More generally, if $T$ is any ordered theory such that each hom-set $T(n,p)$ is a cpo and the composition operation is continuous, then $T$ is an ordered Park theory. For any $f : n+p$, $f^\dagger$ can be given by the formula $f^\dagger = \bigvee_{k \geq 0} f^k \cdot \langle \bot_{n,p}, \mathbf{1}_p \rangle$, where $\bot_{n,p}$ is the least morphism $n \to p$ and $f^k$ is defined by $f^0 = \mathbf{1}_n \oplus 0_p$ and $f^{k+1} = f \cdot \langle f^k, 0_n \oplus \mathbf{1}_p \rangle$, cf. [4, 8].

When $L$ is a complete lattice, both $\mathbf{Mon}_L$ and $\mathbf{Cont}_L$ are semilattice ordered Park theories. Moreover, when $T$ is an ordered theory such that each hom-set is a complete lattice and composition is continuous, then $T$, equipped with the least pre-fixed point operation, is uniquely a semilattice ordered Park theory. In particular, for any set $A$, $\mathbf{Rel}_A$ is semilattice ordered Park theory, where a morphism $n \to p$ is a binary relation from $A \times [n]$ to $A \times [p]$. Composition is relational composition and the distinguished morphism $i_n : 1 \to n$ is the relation $\{(a,(a,i)) : a \in A\}$, for each $i \in [n]$, $n \geq 0$. Several additional examples may be obtained by considering matrix theories $\mathbf{Mat}_S$ over completely idempotent semirings [4] or Blikle nets [3], such as the language semirings $P(A^*)$ of all subsets of $A^*$, where $A$ is any alphabet. In such a theory, a morphism $n \to p$ is an $n \times p$ matrix over $S$, composition is matrix multiplication, etc. When $S$ is the semiring of binary relations over $A$, then $\mathbf{Rel}_A$ is isomorphic to $\mathbf{Mat}_S$.

Further examples include the theory $\mathbf{Lang}_A$ of languages over the alphabet $A$. Here, a morphism $1 \to p$ is a language in $(A \cup [p])^*$. A morphism $n \to p$ is an $n$-tuple of morphisms $1 \to p$. Composition is defined by substitution, so that when $L : 1 \to n$ and $(L_1, \ldots, L_n) : n \to p$, then $L \cdot (L_1, \ldots, L_n) : 1 \to p$ consists of all words that can constructed from the words of $L$ by substituting a word of $L_i$ for each occurrence of the variable $x_i$, for all $i \in [n]$. (Different occurrences of the same variable may be replaced by different words.) A related example is the theory $\mathbf{TreeLang}_\Sigma$ of tree languages over a ranked alphabet $\Sigma$, where $\Sigma$ is any ranked alphabet. In this theory, a morphism $1 \to p$ is a $\Sigma$-tree language in the variables $\{x_1, \ldots, x_p\}$, see also Section 7. A morphism $n \to p$ is an $n$-tuple of tree languages. Composition is OI-substitution, cf. [7]. In both theories $\mathbf{Lang}_A$ and $\mathbf{TreeLang}_\Sigma$, each hom-set is complete lattice and composition is continuous. The subtheory $\mathbf{Reg}_\Sigma$ of $\mathbf{TreeLang}_\Sigma$ determined by just the regular tree languages [15] is also a semilattice ordered Park theory.

A *term* in the language of theories equipped with a dagger operation is a well-formed expression formed from sorted morphism variables (or letters) $f : n \to p$ and the symbols $i_n : 1 \to n$, $i \in [n]$, $n \geq 0$, by the theory operations and $^\dagger$. A term in the language of theories equipped with some additional operations such as $\vee$ or the residuation and star operations introduced later in the sequel may involve those additional operations. Each term $t$ has a source $n$ and a target $p$, noted $t : n \to p$. When the variables $f$ are interpreted as morphisms of appropriate source and target in a theory $T$ equipped with dagger or the other additional operations, each term $t : n \to p$ denotes a morphism $n \to p$ of $T$. When $T$ is ordered, we say that an inequation $t \leq t'$ between terms $t, t' : n \to p$ holds in $T$ if under each interpretation of the variables by morphisms in $T$, the morphism denoted by $t$ is less than the morphism denoted by $t'$ in the ordering of $T$. We say that the equation $t = t'$ holds in $T$ if both $t \leq t'$ and $t' \leq t$ hold. Clearly, when $T$ is semilattice ordered, $t \leq t'$ holds iff $t \vee t' = t'$ does.

**Theorem 3.1** [8, 11] *An inequation $t \leq t'$ between terms in the language of theories equipped with a dagger operation holds in all Park theories iff it holds in all Park theories $\mathbf{Mon}_L$, where $L$ is any complete lattice.*

*An inequation $t \leq t'$ between terms in the language of theories equipped with a dagger operation and an operation $\vee$ holds in all semilattice ordered Park theories iff it holds in all theories $\mathbf{Mon}_L$, where $L$ is any complete lattice.*

**Remark 3.2** *The first part of Theorem 3.1 also holds for the theories $\mathbf{Mon}_P$, $\mathbf{Cont}_P$, or $\mathbf{Cont}_L$, for cpo's $P$ and complete lattices $L$. The second part also holds for the theories $\mathbf{Cont}_L$, equipped with binary supremum, where $L$ is a complete lattice. See [8, 11].*

## 4 Residuated ordered theories

Suppose that $T$ is an ordered theory. Then any morphism $g : p \to q$ in $T$ induces a monotone function $T(n, p) \to T(n, q)$ by right composition: $f \mapsto f \cdot g$, for all $f : n \to p$. When this function has a right adjoint, we have a *Galois connection* that defines a (left) residuation operation.

**Definition 4.1** *Suppose that $T$ is an ordered theory. We call $T$ a **residuated ordered***

**theory** if $T$ is equipped with a binary operation

$$T(n,q) \times T(p,q) \rightarrow T(n,p), \quad n,p,q \geq 0$$
$$(h,g) \mapsto h \Leftarrow g$$

*such that*

$$f \cdot g \leq h \quad \text{iff} \quad f \leq (h \Leftarrow g)$$

*for $f : n \rightarrow p$, $g : p \rightarrow q$ and $h : n \rightarrow q$. We call $h \Leftarrow g$ the* (left) residual of $h$ by $g$.

A **residuated semilattice ordered theory** *is a semilattice ordered theory which is a residuated ordered theory. Morphisms of residuated (semilattice) ordered theories also preserve residuals.*

Note that in a residuated ordered theory, for any $h : n \rightarrow q$ and $g : p \rightarrow q$, $h \Leftarrow g$ is the greatest morphism $f : n \rightarrow p$ with $f \cdot g \leq h$. Thus, an ordered theory can be turned into a residuated ordered theory in at most one way. When $T$ is an ordered theory such that each hom-set $T(n,p)$ is a complete lattice and right composition preserves arbitrary suprema, then $T$ is residuated semilattice ordered theory with $h \Leftarrow g = \bigvee \{f : n \rightarrow p : f \cdot g \leq h\}$ for all $g : p \rightarrow q$ and $h : n \rightarrow q$. In particular, for any complete lattice $L$, $\mathbf{Mon}_L$ and $\mathbf{Cont}_L$ are semilattice ordered residuated theories, as are the theories $\mathbf{Rel}_A$, $\mathbf{Lang}_A$ and $\mathbf{TreeLang}_\Sigma$. Also, the theories $\mathbf{Reg}_\Sigma$ are residuated, since $\mathbf{Reg}_\Sigma$ is closed under the residuation operation of $\mathbf{TreeLang}_\Sigma$, cf. Section 7. When $S$ is a completely idempotent semiring, then $\mathbf{Mat}_S$ is a residuated semilattice ordered theory.

Most of the properties of residuation follow from properties of Galois connections, see e.g., [6], p. 233.

**Proposition 4.2** *Suppose that $T$ is an ordered theory equipped with an operation $\Leftarrow$. Then $T$ is a residuated ordered theory iff (6), (7) hold and $\Leftarrow$ is monotone in its first argument:*

$$h \leq h' \Rightarrow (h \Leftarrow g) \leq (h' \Leftarrow g), \quad h, h' : n \rightarrow q, \ g : p \rightarrow q. \tag{23}$$

*Proof.* Suppose that $T$ is a residuated ordered theory. It is clear that (6) and (7) hold. Suppose now that $g : p \rightarrow q$ and $h, h' : n \rightarrow q$ with $h \leq h'$. Then for all $f : n \rightarrow p$, if $f \cdot g \leq h$ then $f \cdot g \leq h'$. In particular, by (6) we have $(h \Leftarrow g) \cdot g \leq h'$, so that $(h \Leftarrow g) \leq (h' \Leftarrow g)$.

Suppose now that (6), (7) and (23) hold and let $f : n \rightarrow p$, $g : p \rightarrow q$ and $h : n \rightarrow q$. If $f \leq (h \Leftarrow g)$, then $f \cdot g \leq (h \Leftarrow g) \cdot g = h$. And if $f \cdot g \leq h$, then $f \leq ((f \cdot g) \Leftarrow g) \leq (h \Leftarrow g)$.
□

It follows that residuated semilattice ordered theories have an equational axiomatization:

**Corollary 4.3** *Suppose that $T$ is a semilattice ordered theory which is equipped with an operation $\Leftarrow$. Then $T$ is a residuated semilattice ordered theory iff (6), (7) and (8) hold.*

We end this section by presenting some useful properties of residuated ordered theories.

**Proposition 4.4** *Suppose that $T$ is a residuated ordered theory. Then the following hold:*

- If $T(n,p)$ has a least element $\bot_{n,p}$ and $g$ is a morphism $p \to q$, then $\bot_{n,p} \cdot g$ is the least element of $T(n,q)$.

- Suppose that $f_1, f_2 : n \to p$ such that the supremum $f_1 \vee f_2$ exists. Then for any $g : p \to q$, $(f_1 \cdot g) \vee (f_2 \cdot g)$ exists, and $(f_1 \vee f_2) \cdot g = (f_1 \cdot g) \vee (f_2 \cdot g)$.

More generally, if $f_i : n \to p$, $i \in I$ such that $\bigvee_{i \in I} f_i$ exists, where $I$ is any index set, then $\bigvee_{i \in I}(f_i \cdot g)$ also exists, and $(\bigvee_{i \in I} f_i) \cdot g = \bigvee_{i \in I}(f_i \cdot g)$.

*Proof.* Suppose that $f_i : n \to p$, $i \in I$ such that $\bigvee_{i \in I} f_i$ exists, and let $g : p \to q$. Then for any $h : n \to q$,

$$
\begin{aligned}
(\bigvee_{i \in I} f_i) \cdot g \leq h &\Leftrightarrow \bigvee_{i \in I} f_i \leq (h \Leftarrow g) \\
&\Leftrightarrow f_i \leq (h \Leftarrow g) \text{ for all } i \in I \\
&\Leftrightarrow f_i \cdot g \leq h \text{ for all } i \in I.
\end{aligned}
$$

Thus, $h : n \to q$ is an upper bound of $\{f_i \cdot g : i \in I\}$ iff $(\bigvee_{i \in I} f_i) \cdot g \leq h$, showing that $(\bigvee_{i \in I} f_i) \cdot g = \bigvee_{i \in I}(f_i \cdot g)$. $\square$

**Proposition 4.5** *Let $T$ be a residuated ordered theory. Then for all $g, g' : p \to q$ and $h : n \to q$, if $g \leq g'$, then $(h \Leftarrow g') \leq (h \Leftarrow g)$. Thus, residuation is anti-monotone in its second argument.*

*Proof.* Let $g \leq g' : p \to q$ and $h : n \to q$. Since composition is monotone in its second argument, if $f : n \to p$ with $f \cdot g' \leq h$, then $f \cdot g \leq h$. Thus, $(h \Leftarrow g') \leq (h \Leftarrow g)$. $\square$

## 5 Residuated Park theories

In this section, we combine Park theories and residuated (semilattice) ordered theories and complete the proof of Theorem 2.2.

**Lemma 5.1** *Suppose that $T$ is a residuated ordered theory equipped with a dagger operation. If the least pre-fixed point rule (18) holds, then so does (12).*

*Proof.* Assume that the least pre-fixed point rule holds. Then, if $f : n \to n + p$ and $g : n \to p$ with $f \leq (g \Leftarrow \langle g, \mathbf{1}_p \rangle)$, then $f \cdot \langle g, \mathbf{1}_p \rangle \leq g$ and thus $f^\dagger \leq g$. Now (12) follows by taking $f := (g \Leftarrow \langle g, \mathbf{1}_p \rangle)$. $\square$

**Lemma 5.2** *Assume that $T$ is a residuated ordered theory equipped with a dagger operation. If (12) and (21) hold, then the least pre-fixed point rule holds.*

*Proof.* Let $f : n \to n + p$ and $g : n \to p$. If $f \cdot \langle g, \mathbf{1}_p \rangle \leq g$, then $f \leq (g \Leftarrow \langle g, \mathbf{1}_p \rangle)$. Thus, by (12) and (21), $f^\dagger \leq (g \Leftarrow \langle g, \mathbf{1}_p \rangle)^\dagger \leq g$. $\square$

**Definition 5.3** *A **residuated Park theory** is a residuated ordered theory which is a Park theory. A **residuated semilattice ordered Park theory** is a residuated Park theory which is semilattice ordered. Morphisms of residuated (semilattice ordered) Park theories are (semilattice ordered) Park theory morphisms and residuated ordered theory morphisms.*

The theories $\mathbf{Mon}_L$, $\mathbf{Cont}_L$, where $L$ is a complete lattice, are (uniquely) residuated semilattice ordered Park theories, as are the theories $\mathbf{Rel}_A$, $\mathbf{Lang}_A$, $\mathbf{TreeLang}_\Sigma$ and $\mathbf{Reg}_\Sigma$ defined above. When $S$ is a completely idempotent semiring, then $\mathbf{Mat}_S$ is a residuated semilattice ordered Park theory.

Note that unlike semilattice ordered Park theories, residuated semilattice ordered Park theories are defined by equations so that they are closed under quotients. Using the above lemmas, we obtain:

**Theorem 5.4** *Suppose that $T$ is an ordered theory equipped with operations $^\dagger$ and $\Leftarrow$. Then $T$ is a residuated Park theory iff (21), (23) and the inequations (6), (7), (10), (11), (12) hold.*

*Suppose that $T$ is a semilattice ordered theory equipped with operations $^\dagger$ and $\Leftarrow$. Then $T$ is a residuated semilattice ordered Park theory iff the inequations (6) – (12) of Theorem 2.2 hold.*

*Proof.* The first claim follows from Lemma 5.1, Lemma 5.2 and Proposition 4.2. The second claim follows from Lemma 5.1, Lemma 5.2, Proposition 4.4 and Corollary 4.3. □

Thus, a theory equipped with operations $\vee$, $\Leftarrow$ and $^\dagger$ is a semilattice ordered Park theory iff the axioms of Theorem 2.2 hold. The second claim of the following Theorem completes the proof of Theorem 2.2.

**Theorem 5.5** *An inequation between terms in the language of theories equipped with a $^\dagger$ operation holds in all theories $\mathbf{Mon}_L$, where $L$ is a complete lattice, iff it holds in all ordered theories equipped with operations $^\dagger$ and $\Leftarrow$ satisfying (21), (23) and the inequations (6), (7), (10), (11) and (12).*

*An equation between terms in the language of theories equipped with a dagger operation and an operation $\vee$ holds in all theories $\mathbf{Mon}_L$, where $L$ is a complete lattice, iff it holds in all semilattice ordered theories equipped with these operations satisfying (6) – (12).*

*Proof.* Clear from Theorem 3.1 and Theorem 5.4. □

## 6 Dagger vs. star

In matrix theories over certain semirings, including completely idempotent semirings, the dagger operation may be replaced by a star operation, cf. [4]. In this section, we define a (generalized) star operation in all (residuated) semilattice ordered Park theories and study the relationship between dagger and star.

Call a (residuated) semilattice ordered theory $T$ *strict* if for each $n, p$ there exists a least morphism $n \to p$, denoted $\bot_{n,p} : n \to p$, such that $\bot_{n,p} \cdot g = \bot_{n,q}$, for all $g : p \to q$, so that (16) holds. Moreover, call a semilattice ordered theory $T$ *right distributive* if (22) (or (14)) holds. As shown above, every semilattice ordered Park theory is strict with $\bot_{n,p} = (\mathbf{1}_n \oplus 0_p)^\dagger$, every residuated semilattice ordered theory is right distributive, and every residuated semilattice ordered Park theory is both strict and right-distributive.

For any morphism $f : n \to n + p$ in a semilattice ordered theory, let us define

$$f^\tau = f \cdot (\mathbf{1}_n \oplus 0_n \oplus \mathbf{1}_p) \vee (0_n \oplus \mathbf{1}_n \oplus 0_p) : n \to n + n + p.$$

(We assume that $\cdot$ has higher precedence than $\vee$.)

**Definition 6.1** *Let $T$ be a strict (residuated) semilattice ordered theory. If $T$ is equipped with a dagger operation, define a star operation by $f^* = (f^\tau)^\dagger$, for all $f : n \to n + p$. Let $T_*$ denote the resulting theory. If $T$ is equipped with a star operation, define $f^\dagger = f^* \cdot \langle \bot_{n,p}, \mathbf{1}_p \rangle$, for all $f : n \to n + p$. The resulting theory is denoted $T_\dagger$.*

For example, when $T$ is $\mathbf{Rel}_A$, then for any relation $f : n \to n$, $f^*$ is the reflexive-transitive closure of $f$. When $f : n \to n + p$, we may write $f = [f_1, f_2]$, where $f_1 : n \to n$ is the relation $\{((a,i),(b,j)) \in f : j \leq n\}$, and $f_2 : n \to p$ is the relation $\{((a,i),(b,j)) : ((a,i),(b,n+j)) \in f\}$. Then $f^* = [f_1^*, f_1^* f_2]$. Similarly, when $S$ is a completely idempotent semiring and $f : n \to n + p$ in $\mathbf{Mat}_S$, then we may write $f = [f_1, f_2]$ for an $n \times n$ matrix $f_1$ and an $n \times p$ matrix $f_2$. We have that $f^* = [f_1^*, f_1^* f_2]$, where $f_1^* = \sum_{n \geq 0} f_1^n$. In the theories $\mathbf{TreeLang}_\Sigma$ and $\mathbf{Reg}_\Sigma$, the star operation (on morphisms $1 \to 1 + p$) correspond to the $x_i$-iteration operations of [15].

For any term $t$ in the language of theories equipped with operations $\vee$ and $\Leftarrow$, constants $\bot_{n,p}$ and a star operation, we construct a term $t_\dagger$ of the same source and target in the language of theories equipped with operations $\vee$, $\Leftarrow$, constants $\bot_{n,p}$ and a dagger operation by replacing each subterm of the form $s^*$, where $s$ is a term $n \to n + p$, with the term $((s_\dagger)^\tau)^\dagger : n \to n + p$, where

$$(s_\dagger)^\tau = s_\dagger \cdot (\mathbf{1}_n \oplus 0_n \oplus \mathbf{1}_p) \vee (0_n \oplus \mathbf{1}_n \oplus 0_p) : n \to n + n + p.$$

Conversely, any term $t$ in the language theories equipped with operations $\vee$ and $\Leftarrow$, constants $\bot_{n,p}$ and a dagger operation may be transformed into a term $t_*$ in the language of theories equipped with operations $\vee$, $\Leftarrow$, constants $\bot_{n,p}$ and a star operation by replacing each subterm of the form $s^\dagger : n \to p$, where $s : n \to n+p$, with $(s_*)^* \cdot \langle \bot_{n,p}, \mathbf{1}_p \rangle : n \to p$.

**Theorem 6.2** *Let $T$ denote a strict right distributive (residuated) semilattice ordered theory. Suppose that $T$ is equipped with a dagger operation which satisfies the parameter equation (20). Then the* star parameter equation

$$f^* \cdot (\mathbf{1}_n \oplus g) = (f \cdot (\mathbf{1}_n \oplus g))^*, \quad f : n \to n + p, \ g : p \to q \tag{24}$$

*and the equation*

$$f^* = (f^\tau)^* \cdot \langle \bot_{n,n+p}, \mathbf{1}_{n+p} \rangle, \quad f : n \to n + p \tag{25}$$

*hold in $T_*$. Moreover, $t = t_{*\dagger}$ holds in $T$ for all terms $t$ in the language of theories equipped with operations $\vee$, $\Leftarrow$, constants $\bot_{n,p}$ and a dagger operation.*

*Suppose that $T$ is equipped with a star operation. If the star parameter equation (24) and (25) hold in $T$, then the parameter equation (20) holds in $T_\dagger$ for the dagger operation. Moreover, $t = t_{\dagger *}$ holds in $T$ for all terms $t$ in the language of theories equipped with operations $\vee$, $\Leftarrow$, constants $\bot_{n,p}$ and a star operation.*

The proof of this result is given in the Appendix. By Theorem 6.2, under certain assumptions, properties of the star operation are reflected by corresponding properties of dagger, and vice versa. But sometimes there exist stronger results.

The *star fixed point inequation* is

$$f \cdot \langle f^*, 0_n \oplus 1_p \rangle \vee (1_n \oplus 0_p) \leq f^* \qquad (26)$$

where $f : n \to n + p$. The *star least pre-fixed point rule* is

$$f \cdot \langle g, 0_n \oplus 1_p \rangle \vee h \leq g \quad \Rightarrow \quad f^* \cdot \langle h, 0_n \oplus 1_p \rangle \leq g, \qquad (27)$$

where $f, g, h : n \to n + p$.

In the next lemma and the subsequent propositions, we let $T$ denote a strict right distributive semilattice ordered theory.

**Lemma 6.3** *Suppose that $T$ is equipped with a star operation. Then the star least pre-fixed point rule (27) holds iff*

$$f \cdot \langle g, 0_n \oplus 1_p \rangle \leq g \quad \Rightarrow \quad f^* \cdot \langle g, 0_n \oplus 1_p \rangle \leq g \qquad (28)$$

*for all $f, g : n \to n + p$.*

*Proof.* Suppose first that (28) holds. If $f \cdot \langle g, 0_n \oplus 1_p \rangle \vee h \leq g$, where $f, g, h : n \to n + p$, then $f \cdot \langle g, 0_n \oplus 1_p \rangle \leq g$ and $h \leq g$. Thus, $f^* \cdot \langle h, 0_n \oplus 1_p \rangle \leq f^* \cdot \langle g, 0_n \oplus 1_p \rangle \leq g$.

Suppose now that the star least pre-fixed point rule (27) holds. If $f \cdot \langle g, 0_n \oplus 1_p \rangle \leq g$, then $f \cdot \langle g, 0_n \oplus 1_p \rangle \vee g \leq g$. Thus, $f^* \cdot \langle g, 0_n \oplus 1_p \rangle \leq g$. □

**Proposition 6.4** *Suppose that $T$ is equipped with a dagger operation. If the fixed point inequation (10) holds in $T$, then the star fixed point inequation (26) holds in $T_*$. If the parameter inequation (11) and least pre-fixed point rule (18) hold in $T$, then the star least pre-fixed point rule (27) holds in $T_*$.*

*Proof.* Suppose first that (10) holds in $T$ and let $f : n \to n + p$. Then, using (14),

$$\begin{aligned}
f \cdot \langle f^*, 0_n \oplus 1_p \rangle \vee (1_n \oplus 0_p) &= f \cdot \langle (f^\tau)^\dagger, 0_n \oplus 1_p \rangle \vee (1_n \oplus 0_p) \\
&= (f \cdot (1_n \oplus 0_n \oplus 1_p) \vee (0_n \oplus 1_n \oplus 0_p)) \cdot \langle (f^\tau)^\dagger, 1_{n+p} \rangle \\
&= f^\tau \cdot \langle (f^\tau)^\dagger, 1_{n+p} \rangle \\
&\leq (f^\tau)^\dagger = f^*,
\end{aligned}$$

proving that (26) holds in $T_*$.

Suppose now that (11) and (18) hold in $T$, and let $f, g : n \to n + p$ with

$$f \cdot \langle g, 0_n \oplus 1_p \rangle \leq g.$$

Since

$$\begin{aligned}
f^\tau \cdot (1_n \oplus \langle g, 0_n \oplus 1_p \rangle) \cdot \langle g, 1_{n+p} \rangle &= f^\tau \cdot \langle g, g, 0_n \oplus 1_p \rangle \\
&= (f \cdot (1_n \oplus 0_n \oplus 1_p) \vee (0_n \oplus 1_n \oplus 0_p)) \cdot \langle g, g, 0_n \oplus 1_p \rangle \\
&= f \cdot \langle g, 0_n \oplus 1_p \rangle \vee g \\
&\leq g,
\end{aligned}$$

by (27) we have

$$(f^\tau \cdot (1_n \oplus \langle g, 0_n \oplus 1_p \rangle))^\dagger \leq g.$$

But by the parameter inequation (11),

$$f^* \cdot \langle g, 0_n \oplus \mathbf{1}_p \rangle = (f^\tau)^\dagger \cdot \langle g, 0_n \oplus \mathbf{1}_p \rangle \leq f^\tau \cdot (\mathbf{1}_n \oplus \langle g, 0_n \oplus \mathbf{1}_p \rangle))^\dagger$$

and thus $f^* \cdot \langle g, 0_n \oplus \mathbf{1}_p \rangle \leq g$. □

**Proposition 6.5** *Suppose that $T$ is equipped with a star operation. If the star fixed point inequation (26) holds in $T$, then the fixed point inequation (10) holds in $T_\dagger$. And if the star least pre-fixed point rule (27) holds in $T$, then the least pre-fixed point rule (18) holds in $T_\dagger$.*

*Proof.* Suppose first that (26) holds in $T$ and let $f : n \to n + p$. Then

$$\begin{aligned} f \cdot \langle f^\dagger, \mathbf{1}_p \rangle &= f \cdot \langle f^* \cdot \langle \bot_{n,p}, \mathbf{1}_p \rangle, \mathbf{1}_p \rangle \\ &= f \cdot \langle f^* \cdot \langle \bot_{n,p}, \mathbf{1}_p \rangle, \mathbf{1}_p \rangle \vee \bot_{n,p} \\ &= (f \cdot \langle f^*, 0_n \oplus \mathbf{1}_p \rangle \vee (\mathbf{1}_n \oplus 0_p)) \cdot \langle \bot_{n,p}, \mathbf{1}_p \rangle \\ &\leq f^* \cdot \langle \bot_{n,p}, \mathbf{1}_p \rangle = f^\dagger. \end{aligned}$$

Suppose now that (27) holds in $T$, and let $f : n \to n+p$ and $g : n \to p$ with $f \cdot \langle g, \mathbf{1}_p \rangle \leq g$. Then

$$f \cdot \langle 0_n \oplus g, 0_n \oplus \mathbf{1}_p \rangle = 0_n \oplus f \cdot \langle g, \mathbf{1}_p \rangle \leq 0_n \oplus g$$

and thus $f^* \cdot \langle 0_n \oplus g, 0_n \oplus \mathbf{1}_p \rangle \leq 0_n \oplus g$, by Lemma 6.3. Composing both sides with $\langle \bot_{n,p}, \mathbf{1}_p \rangle$ this gives $f^* \cdot \langle g, \mathbf{1}_p \rangle \leq g$, We conclude that

$$f^\dagger = f^* \cdot \langle \bot_{n,p}, \mathbf{1}_p \rangle \leq f^* \cdot \langle g, \mathbf{1}_p \rangle \leq g. \quad \Box$$

Propositions 6.4 and 6.5 refine results from [13].

**Corollary 6.6** *Suppose that $T$ is a strict right distributive semilattice ordered theory equipped with a dagger operation. Then $T$ is a semilattice ordered Park theory iff $T_*$ satisfies (24), (25), (26) and (27).*

*Suppose that $T$ is a strict right distributive semilattice ordered theory equipped with a star operation. Then $T_\dagger$ is a semilattice ordered Park theory iff $T$ satisfies (24), (25), (26) and (27).*

Of course, (24) may be replaced by

$$f^* \cdot (\mathbf{1}_n \oplus g) \leq (f \cdot (\mathbf{1}_n \oplus g))^*, \quad f : n \to n + p, \; g : p \to q. \tag{29}$$

The above Corollary motivates the following definition, see also [13].

**Definition 6.7** *A **star Park theory** is a strict right distributive semilattice ordered theory equipped with a star operation satisfying (24), (25), (26) and (27).*

It is clear that the star operation is monotone in any star Park theory:

$$f^* \leq (f \vee g)^*, \quad f, g : n \to n + p. \tag{30}$$

**Corollary 6.8** *If $T$ is a semilattice ordered Park theory, then $T_*$ is a star Park theory. If $T$ is a star Park theory, then $T_\dagger$ is a (semilattice ordered) Park theory. Moreover, we have $T = T_{*\dagger}$ for all semilattice ordered Park theories $T$, and $T = T_{\dagger*}$ for all star Park theories $T$.*

*Proof.* The first two claims are clear from the above results. Suppose that $T$ is a semilattice ordered Park theory. The fact that $T = T_{*\dagger}$ follows from the fact that the equation $f^\dagger = (f^\dagger)_{*\dagger}$ holds in $T$, where $f : n \to n + p$. Finally, when $T$ is a star Park theory, then since $f^* = (f^*)_{\dagger*}$ holds in $T$, for all $f : n \to n + p$, we have that $T = T_{\dagger*}$. □

**Remark 6.9** *By introducing appropriate morphisms between star Park theories, it follows that the category of semilattice ordered Park theories is isomorphic to the category of star Park theories by functors that preserve morphisms.*

We now combine residuation with the star operation.

**Proposition 6.10** *Let $T$ denote a strict residuated semilattice ordered theory. Suppose that $T$ is equipped with a monotone star operation. Then (28) holds iff*

$$(g{\Leftarrow}\langle g, 0_n \oplus \mathbf{1}_p\rangle)^* \ \leq \ (g{\Leftarrow}\langle g, 0_n \oplus \mathbf{1}_p\rangle) \tag{31}$$

*for all $g : n \to n + p$.*

*Proof.* Suppose that (31) holds. If $f \cdot \langle g, 0_n \oplus \mathbf{1}_p\rangle \leq g$, where $f, g : n \to n + p$, then $f \leq (g{\Leftarrow}\langle g, 0_n \oplus \mathbf{1}_p\rangle)$ so that $f^* \leq (g{\Leftarrow}\langle g, 0_n \oplus \mathbf{1}_p\rangle)^* \leq (g{\Leftarrow}\langle g, 0_n \oplus \mathbf{1}_p\rangle)$ yielding $f^* \cdot \langle g, 0_n \oplus \mathbf{1}_p\rangle \leq g$.

Suppose now that (28) holds and let $g : n \to n + p$. Since $(g{\Leftarrow}\langle g, 0_n \oplus \mathbf{1}_p\rangle) \cdot \langle g, 0_n \oplus \mathbf{1}_p\rangle \leq g$, thus $(g{\Leftarrow}\langle g, 0_n \oplus \mathbf{1}_p\rangle)^* \cdot \langle g, 0_n \oplus \mathbf{1}_p\rangle \leq g$, so that $(g{\Leftarrow}\langle g, 0_n \oplus \mathbf{1}_p\rangle)^* \leq (g{\Leftarrow}\langle g, 0_n \oplus \mathbf{1}_p\rangle)$. □

We summarize the results of this section. Recall that every residuated semilattice ordered theory is right distributive.

**Definition 6.11** *A **residuated star Park theory** is a strict residuated semilattice ordered theory which is a star Park theory.*

**Theorem 6.12** *Suppose that $T$ is a strict residuated semilattice ordered theory equipped with a star operation. Then $T$ is a residuated star Park theory iff $T$ satisfies (25), (26), (29), (30) and (31).*

*Proof.* Suppose first that $T$ is a residuated star Park theory. Then (25) and (26) hold by definition, and since (24) holds, so does (29). Moreover, it is clear that star is monotone, so that (30) holds. Thus, by Proposition 6.10, (31) also holds.

Suppose now that (25), (26), (31), (29) and (30) hold. Then by Proposition 6.10 and Lemma 6.3, the star least pre-fixed point rule holds. It follows now that $(f \cdot (\mathbf{1}_n \oplus g))^* \leq f^* \cdot (\mathbf{1}_n \oplus g)$ for all $f : n \to n + p$ and $g : p \to q$, so that (24) holds by (29). □

**Theorem 6.13** *An (in)equation between terms in the language of theories equipped with operations $\vee$ and $*$ and constants $\bot_{n,p}$ holds in all theories $\mathbf{Mon}_L$, where $L$ is any complete lattice iff it holds in all strict residuated semilattice ordered theories equipped with a star operation satisfying the axioms of Theorem 6.12.*

*Proof.* This follows from Theorem 6.12, Theorem 6.2 and Theorem 2.2. $\square$

We end this section by presenting an interesting property of the star operation in residuated star Park theories.

**Proposition 6.14** *Suppose that $T$ is a semilattice ordered theory equipped with a star operation. If the star fixed point inequation (26) holds in $T$, then*

$$(\mathbf{1}_n \oplus 0_p) \vee f \vee f^* \cdot \langle f^*, 0_n \oplus \mathbf{1}_p \rangle \leq f^*,$$

*for all $f : n \to n+p$. If both the star fixed point inequation (26) and the star least pre-fixed point rule (27) hold in $T$, then*

$$(\mathbf{1}_n \oplus 0_p) \vee f \vee g \cdot \langle g, 0_n \oplus \mathbf{1}_p \rangle \leq g \;\Rightarrow\; f^* \leq g, \tag{32}$$

*for all $f, g : n \to n+p$.*

*Proof.* To prove the first claim, suppose that the star fixed point inequation holds in $T$ and let $f : n \to n+p$ in $T$. The fact that $\mathbf{1}_n \oplus 0_p \leq f^*$ is immediate. Since this holds and the theory operations are monotone, also

$$f = f \cdot \mathbf{1}_n = f \cdot \langle \mathbf{1}_n \oplus 0_p, 0_n \oplus \mathbf{1}_p \rangle \leq f \cdot \langle f^*, 0_n \oplus \mathbf{1}_p \rangle \leq f^*.$$

Finally, since $f \leq f^*$ and the theory operations are monotone, from $f \cdot \langle f^*, 0_n \oplus \mathbf{1}_p \rangle \leq f^*$ we obtain that $f^* \cdot \langle f^*, 0_n \oplus \mathbf{1}_p \rangle \leq f^*$.

Suppose now that the star least pre-fixed point rule also holds. Then $*$ is monotone. Suppose that $(\mathbf{1}_n \oplus 0_p) \vee f \vee g \cdot \langle g, 0_n \oplus \mathbf{1}_p \rangle \leq g$, where $f, g : n \to n+p$. Then $g \cdot \langle g, 0_n \oplus \mathbf{1}_p \rangle \leq g$ and thus $g^* \cdot \langle g, 0_n \oplus \mathbf{1}_p \rangle \leq g$. But $f \leq g$ and $\mathbf{1}_n \oplus 0_p \leq g$, so that $f^* \leq g^*$ and

$$f^* = f^* \cdot \langle \mathbf{1}_n \oplus 0_p, 0_n \oplus \mathbf{1}_p \rangle \leq g^* \cdot \langle g, 0_n \oplus \mathbf{1}_p \rangle \leq g. \quad \square$$

**Corollary 6.15** *In a star Park theory, for each morphism $f : n \to n+p$, $f^*$ is the least morphism $g : n \to n+p$ with*

$$(\mathbf{1}_n \oplus 0_p) \vee f \vee g \cdot \langle g, 0_n \oplus \mathbf{1}_p \rangle \leq g. \tag{33}$$

**Proposition 6.16** *Suppose that $T$ is a strict residuated semilattice ordered theory equipped with a monotone star operation. If (31) and (32) hold, then so does the star least pre-fixed point rule (27).*

*Proof.* Let $f, g : n \to n+p$ in $T$ with $f \cdot \langle g, 0_n \oplus \mathbf{1}_p \rangle \leq g$. Then $f \leq (g \Leftarrow \langle g, 0_n \oplus \mathbf{1}_p \rangle)$ and thus $f^* \leq (g \Leftarrow \langle g, 0_n \oplus \mathbf{1}_p \rangle)^* \leq (g \Leftarrow \langle g, 0_n \oplus \mathbf{1}_p \rangle)$, so that $f^* \cdot \langle g, 0_n \oplus \mathbf{1}_p \rangle \leq g$. $\square$

**Corollary 6.17** *Suppose that $T$ is a strict residuated semilattice ordered theory equipped with a star operation. Then $T$ is a residuated star Park theory iff for each $f : n \to n+p$, $f^*$ is the least morphism $g : n \to n+p$ such that (33) holds.*

## 6.1 Scalar dagger and star

It is known, cf. [2, 5, 4], that in Park theories, the dagger operation is determined by its restriction to *scalar* morphisms $1 \to 1 + p$, since

$$\langle f, g \rangle^\dagger = \langle f^\dagger \cdot \langle h^\dagger, \mathbf{1}_p \rangle, h^\dagger \rangle$$

with

$$h = g \cdot \langle f^\dagger, \mathbf{1}_{m+p} \rangle$$

for all $f : n \to n + m + p$ and $g : m \to n + m + p$. Moreover, if an ordered theory $T$ is equipped with a scalar dagger operation $f \mapsto f^\dagger$, $f : 1 \to 1 + p$, satisfying the scalar fixed point inequation

$$f \cdot \langle f^\dagger, \mathbf{1}_p \rangle \leq f^\dagger, \quad f : 1 \to 1 + p,$$

the scalar parameter inequation

$$f^\dagger \cdot g \leq (f \cdot (\mathbf{1}_n \oplus g))^\dagger, \quad f : 1 \to 1 + p, \ g : p \to q,$$

and the scalar least pre-fixed point rule

$$f \cdot \langle g, \mathbf{1}_p \rangle \leq g \ \Rightarrow \ f^\dagger \leq g, \quad f : 1 \to 1 + p, \ g : 1 \to p$$

then there is a unique Park theory structure on $T$ extending the scalar dagger operation.

A similar fact holds for star Park theories, cf. [13]. If $T$ is a star Park theory, then the star operation is determined by its restriction to scalar morphisms $1 \to 1 + p$. And if $T$ is a strict semilattice ordered theory equipped with a scalar star operation, satsifying the "scalar versions" of the axioms star Park theories, then the star operation can be continued in a unique way to all morphisms $n \to n + p$ such that $T$ becomes a star Park theory.

Also, if $T$ is a residuated ordered theory, then the residuation operation $\Leftarrow$ is uniquely determined by its restriction to scalar morphisms, since

$$\langle h, h' \rangle \!\Leftarrow\! g = \langle h \!\Leftarrow\! g, h' \!\Leftarrow\! g \rangle$$

for all $h : n \to q$, $h' : m \to q$ and $g : p \to q$. And if $T$ is an ordered theory with an operation $h \!\Leftarrow\! g$, defined for morphisms $h : 1 \to q$ and $g : p \to q$, such that $f \cdot g \leq h$ iff $f \leq (h \!\Leftarrow\! g)$ holds for all $f : 1 \to p$, then $T$ is uniquely a residuated ordered theory.

In conclusion, in all of our results, we may replace the dagger, star and residuation operations by their scalar versions, and in fact we may replace theories by clones.

# 7 Regular tree languages

In this section, we present an application to regular tree languages. When $\Sigma$ is ranked alphabet and $n \geq 0$, we denote by $T_\Sigma(X_n)$ the set of all (finite) $\Sigma$-trees in the variables $x_1, \ldots, x_n$, cf. [15]. We recall that a ($\Sigma$-)tree language is a subset of $T_\Sigma(X_n)$, for some $n \geq 0$. As mentioned above, the theory **TreeLang**$_\Sigma$ whose morphisms $n \to p$ are $n$-tuples of tree languages in $T_\Sigma(X_p)$ is a semilattice ordered Park theory containing the theory **Reg**$_\Sigma$ of regular languages as a subtheory. Actually, **Reg**$_\Sigma$ is the least semilattice ordered Park subtheory of **TreeLang**$_\Sigma$ containing the finite languages.

**Proposition 7.1** **TreeLang**$_\Sigma$ *and* **Reg**$_\Sigma$ *are residuated semilattice ordered Park theories.*

*Proof.* When $L : 1 \to q$ and $K : p \to q$ in **TreeLang**$_\Sigma$, define $L/K = \{t \in T_\Sigma(X_p) : \{t\} \cdot K \cap L \neq \emptyset\} : 1 \to p$. Using this quotient operation, we have for $L$ and $K$ as above that $(L \Leftarrow K) = \overline{L/\overline{K}}$, where $\overline{K}$ denotes the component-wise complement of $K$ with respect to $T_\Sigma(X_q)$ and $\overline{L/\overline{K}}$ is the complement of $L/\overline{K}$ with respect to $T_\Sigma(X_p)$. Since regular languages are closed under taking quotients and complements, if $L$ is regular, then so is $L \Leftarrow K$. $\square$

**Definition 7.2** *Suppose that $T$ is a strict semilattice ordered theory. We call a morphism $f : 1 \to p$ strict if*

$$f \cdot \langle 1_p, \ldots, (i-1)_p, \bot_{1,p}, (i+1)_p, \ldots, p_p \rangle = \bot_{1,p}$$

*for all $i \in [p]$. Moreover, we call $f$ distributive if*

$$f \cdot \langle 1_{p+1}, \ldots, (i-1)_{p+1}, i_{p+1} \vee (i+1)_{p+1}, (i+2)_{p+1}, \ldots, (p+1)_{p+1} \rangle$$
$$= f \cdot \langle 1_{p+1}, \ldots, (i-1)_{p+1}, i_{p+1}, (i+2)_{p+1}, \ldots, (p+1)_{p+1} \rangle$$
$$\vee f \cdot \langle 1_{p+1}, \ldots, (i-1)_{p+1}, (i+1)_{p+1}, (i+2)_{p+1}, \ldots, (p+1)_{p+1} \rangle$$

*holds for all $i \in [p]$.*

For example, for each letter $\sigma \in \Sigma_p$, the morphism $\{\sigma(x_1, \ldots, x_p)\} : 1 \to p$ is strict and distributive. More generally, a tree language $L : 1 \to p$ is strict and distributive iff each tree $t \in L$ contains exactly one occurrence of each variable $x_i \in X_p$.

The following theorem is a reformulation of a result from [9, 12]. For any term $t$ over $\Sigma$ in the language of theories equipped with operations $\vee$ and $^\dagger$, we let $|t|$ denote the morphism denoted by $t$ in **Reg**$_\Sigma$ when each letter $\sigma \in \Sigma_p$ is interpreted as the language $\{\sigma(x_1, \ldots, x_p)\} : 1 \to p$.

**Theorem 7.3** *Suppose that $t$ and $t'$ are terms $n \to p$ over $\Sigma$ in the language of theories equipped with operations $\vee$ and $^\dagger$. Then $|t| = |t'|$ iff $t = t'$ holds in all semilattice ordered Park theories under all interpretations of the letters in $\Sigma$ by strict distributive morphisms.*

**Corollary 7.4** *Suppose that $t$ and $t'$ are terms $n \to p$ over $\Sigma$ in the language of theories equipped with operations $\vee$ and $^\dagger$. Then $|t| = |t'|$ iff $t = t'$ holds in all residuated semilattice ordered Park theories under all interpretations of the letters in $\Sigma$ by strict distributive morphisms.*

*Proof.* If $|t| = |t'|$, then by Theorem 7.3, $t = t'$ holds in all semilattice ordered Park theories under all interpretations of the letters in $\Sigma$ by strict distributive morphisms and thus in all residuated semilattice ordered Park theories under all such interpretations. In order to prove the other direction, suppose that $t = t'$ holds in all residuated semilattice ordered Park theories under all interpretations of the letters in $\Sigma$ by strict distributive morphisms. Then in particular $t = t'$ holds in **Reg**$_\Sigma$ when each letter $\sigma \in \Sigma_p$ is interpreted as the morphism $\{\sigma(x_1, \ldots, x_p)\}$. This means that $|t| = |t'|$. $\square$

Thus, terms $t : 1 \to p$ and $t' : 1 \to p$ over $\Sigma$ denote the same regular tree language iff the equation $t = t'$ can be proved from the (equational) axioms of residuated Park theories and the equations expressing that each $\sigma \in \Sigma$ is strict and distributive. A similar result holds for terms in the language of theories equipped with operations $\vee$ and $^*$ and constants $\perp_{n,p}$.

# Appendix

In this appendix, we prove Theorem 6.2.

Suppose that $T$ is equipped with a dagger operation satisfying the parameter equation. Then for all $f : n \to n+p$ and $g : p \to q$,

$$\begin{aligned}
(f \cdot (\mathbf{1}_n \oplus g))^* &= (f \cdot (\mathbf{1}_n \oplus g) \cdot (\mathbf{1}_n \oplus 0_n \oplus \mathbf{1}_q) \vee (0_n \oplus \mathbf{1}_n \oplus 0_q))^\dagger \\
&= (f \cdot (\mathbf{1}_n \oplus 0_n \oplus g) \vee (0_n \oplus \mathbf{1}_n \oplus 0_q))^\dagger \\
&= ((f \cdot (\mathbf{1}_n \oplus 0_n \oplus \mathbf{1}_p) \vee (0_n \oplus \mathbf{1}_n \oplus 0_p)) \cdot (\mathbf{1}_n \oplus \mathbf{1}_n \oplus g))^\dagger \\
&= (f \cdot (\mathbf{1}_n \oplus 0_n \oplus \mathbf{1}_p) \vee (0_n \oplus \mathbf{1}_n \oplus 0_p))^\dagger \cdot (\mathbf{1}_n \oplus g) \\
&= f^* \cdot (\mathbf{1}_n \oplus g),
\end{aligned}$$

where we have used the assumption that $T$ is right distributive. It is easy to see that

$$f^{\tau\tau} = f \cdot (\mathbf{1}_n \oplus 0_{2n} \oplus \mathbf{1}_p) \vee (0_{2n} \oplus \mathbf{1}_n \oplus 0_p) \vee (0_n \oplus \mathbf{1}_n \oplus 0_{n+p})$$

and thus

$$\begin{aligned}
f^{\tau\tau} \cdot (\mathbf{1}_n \oplus \langle \bot_{n,n+p}, \mathbf{1}_{n+p} \rangle)) &= f \cdot (\mathbf{1}_n \oplus 0_n \oplus \mathbf{1}_p) \vee (0_n \oplus \mathbf{1}_n \oplus 0_p) \vee \bot_{n,n+n+p} \\
&= f^\tau,
\end{aligned}$$

since by assumption $\bot_{n,n+n+p}$ is the least morphism $n \to n+n+p$. Thus,

$$\begin{aligned}
(f^\tau)^* \cdot \langle \bot_{n,n+p}, \mathbf{1}_{n+p} \rangle &= (f^{\tau\tau})^\dagger \cdot \langle \bot_{n,n+p}, \mathbf{1}_{n+p} \rangle \\
&= (f^{\tau\tau} \cdot (\mathbf{1}_n \oplus \langle \bot_{n,n+p}, \mathbf{1}_{n+p} \rangle))^\dagger \\
&= (f^\tau)^\dagger = f^*.
\end{aligned}$$

Let $t$ be any term in the language of theories equipped with operations $\vee$, $\Leftarrow$, constants $\bot_{n,p}$ and a dagger operation. The fact that $t = t_{*\dagger}$ holds in $T$ is proved by induction on the structure of $t$. The only nontrivial case is when $t = s^\dagger$, where $s$ is a term $n \to n+p$. But then, using the induction hypothesis in the 4th line, the parameter equation in the 5th line, and the assumption that $T$ is strict and right distributive in 7th, 8th and 9th lines, we have that

$$\begin{aligned}
t_{*\dagger} &= (s^\dagger)_{*\dagger} \\
&= ((s_*)^* \cdot \langle \bot_{n,p}, \mathbf{1}_p \rangle)_\dagger \\
&= ((s_*)^*)_\dagger \cdot \langle \bot_{n,p}, \mathbf{1}_p \rangle \\
&= ((s_{*\dagger})^\tau)^\dagger \cdot \langle \bot_{n,p}, \mathbf{1}_p \rangle \\
&= (s^\tau)^\dagger \cdot \langle \bot_{n,p}, \mathbf{1}_p \rangle \\
&= (s^\tau \cdot (\mathbf{1}_n \oplus \langle \bot_{n,p}, \mathbf{1}_p \rangle))^\dagger \\
&= ((s \cdot (\mathbf{1}_n \oplus 0_p \oplus \mathbf{1}_p) \vee (0_n \oplus \mathbf{1}_n \oplus 0_p)) \cdot (\mathbf{1}_n \oplus \langle \bot_{n,p}, \mathbf{1}_p \rangle))^\dagger \\
&= (s \vee (0_n \oplus \bot_{n,p}))^\dagger \\
&= (s \vee \bot_{n,n+p})^\dagger \\
&= s^\dagger \\
&= t.
\end{aligned}$$

Suppose now that $T$ is equipped with a star operation which satisfies the star parameter equation. To prove that the parameter equation holds in $T_\dagger$ for the dagger operation, let $f : n \to n+p$ and $g : p \to q$. Then, using (16) and the assumption that $T$ is strict,

$$\begin{aligned}
(f \cdot (\mathbf{1}_n \oplus g))^\dagger &= (f \cdot (\mathbf{1}_n \oplus g))^* \cdot \langle \bot_{n,q}, \mathbf{1}_q \rangle \\
&= f^* \cdot (\mathbf{1}_n \oplus g) \cdot \langle \bot_{n,q}, \mathbf{1}_q \rangle \\
&= f^* \cdot \langle \bot_{n,q}, g \rangle \\
&= f^* \cdot \langle \bot_{n,p}, \mathbf{1}_p \rangle \cdot g \\
&= f^\dagger \cdot g,
\end{aligned}$$

proving that the parameter equation holds in $T_\dagger$. Assume now that (25) also holds. We prove that $t = t_{\dagger*}$ holds in $T$, for all terms $t$ in the language of theories equipped with operations $\vee$, $\Leftarrow$, constants $\bot_{n,p}$ and star.

The proof goes by induction on the structure of $t$. The only nontrivial case is when $t = s^*$. In this case, using (25),

$$\begin{aligned}
t_{\dagger*} &= (s^*)_{\dagger*} \\
&= (((s_\dagger)^\tau)^\dagger)_* \\
&= (((s_\dagger)^\tau)_*)^* \cdot \langle \bot_{n,n+p}, \mathbf{1}_{n+p} \rangle \\
&= ((s_{\dagger*})^\tau)^* \cdot \langle \bot_{n,n+p}, \mathbf{1}_{n+p} \rangle \\
&= (s^\tau)^* \cdot \langle \bot_{n,n+p}, \mathbf{1}_{n+p} \rangle \\
&= s^* \\
&= t. \quad \square
\end{aligned}$$